\def\Ref#1{\thinspace[\thinspace #1\thinspace]}

\def\GeVc2{\hbox{GeV$\!/c^2$}}

\def\frac#1#2{{#1\over#2}}
\def\n{\hfill\break}
\nopagenumbers
\footline={\ifnum\pageno>0 \hfil\folio\hfil \else \hfil \fi}

\pageno=0
\null\vskip -21pt
\line{{lp01jjm.tex\hfill dias-stp-01-09}}
\vskip 2ex
\centerline{\bf Jets signal for Higgs particle detection at LHC}
\vskip 6ex
\centerline{T. Garavagli,$^{*}$} 
\vskip 2ex
\centerline{\it Institi\'uid \'Ard-l\'einn Bhaile \'Atha Cliath, Baile
\'Atha Cliath  4, \'Eire$^{\dagger}$}

\vskip42pt
\vbox{ \leftskip11pt \rightskip15pt

A method using jets is investigated for detecting the Higgs boson at LHC in the
mass range about $114$ \GeVc2, suggested by LEP experiments. Higgs bosons are  produced
in association with a $t \bar{t}$ pair, and both $t$ and $\bar{t}$ 
decay  semileptonically to reduce the QCD background. 
After  appropriate cuts, the signal is
compared with the main background, $t \bar{t} + 2$ jets.
This estimate, using a reasonable approximation for the dominant
background $t\bar {t}gg$, suggests a $5.1\sigma$ effect. 
This method is seen to be complimentary to the  two gamma signal. 
The $t\bar{t}Z$ channel, with $Z$ decaying to $l^+ l^-$, may be used to
reduce theoretical uncertainties in determining the $t \bar{t}H$ signal.  

\vskip 10pt
XX International Symposium on Lepton and Photon Interactions at High
Energies, Rome 2001. Abstract Number 104.
} 
\vfil
\hrule width 1in \vskip1ex
\item{$^*$}{Email address: bronco@stp.dias.ie }
\item{$^\dagger$}{Also {\it Institi\'uid Teicneola\'iochta Bhaile 
\'Atha Cliath.}}
\eject

\noindent{\bf 1. INTRODUCTION}
\n
\par
The Standard Model Higgs with mass up to 114 \GeVc2\ has been 
suggested by fits to electroweak parameters\Ref{1} and recent  experiments at 
LEP \Ref{2--3}. 
These results could be confirmed at the LHC where it 
should be possible  to find a Higgs boson with mass near 114
\GeVc2. 
For this mass, the rare decay $H \to
\gamma\gamma$ was suggested\Ref{4}, where the Higgs boson $H$ is
produced in association with either a $W^\pm$ boson or a $t \bar{t}$ pair
in $pp$ scattering experiments.  A charged hard lepton from $W^\pm$ or one
of the $t$ quarks is required as a trigger in order to reduce the QCD
background. However, since the $2\gamma$ mode is a rare decay, it may be
difficult to extract this signal from the complicated background.
I study the possibility of searching for an intermediate Higgs
boson in the two-jet channel by reconstructing its invariant mass. I
consider Higgs bosons that are produced in association with $t \bar{t}$
pairs only.  This allows both the $t$ and $\bar{t}$ to be tagged using
semileptonic decays to cut down on the QCD background.
Similar method have been used in top searches
at the TEVATRON. Although this method had been suggested for the
SSC\Ref{5},
there has been considerable improvement in b vertex recognition
and NLO QCD corrections to the $g\,g\rightarrow H$ cross section\Ref{6}
since then, and the
probable success of this search method at the LHC has increased.
\n\n
\noindent{\bf 2. SIGNAL AND BACKGROUND}
\n
\par
The main process of interest is 
$$
  pp\to t\bar{t} H + X,  \eqno(1)
$$
with
$$\eqalignno{
  t       &\to b + (W^+\to l^+ +\nu),                &(2) \cr
  \bar{t} &\to \bar{b} + (W^-\to l^- +\bar{\nu}),    &(3) \cr
  H       &\to b\bar{b} \quad{\rm or}\quad c\bar{c}, &(4) \cr}
$$
where $l$ stands for either an electron or a muon. The final signal searched
for is 
$$
  {\rm 2~leptons + 4~jets + missing}~E_T.  \eqno(5)
$$
It has been know for some time \Ref{4} and \Ref{7} that $t\bar{t}H$ production at $pp$
colliders can be very well approximated by two-gluon fusion alone. The 
tree level cross section as a function of the Higgs mass is found using the
method of 
Kunzst \Ref{4} for $m_t =175~\GeVc2$. 
As an example, I take $m_H
=114~\GeVc2$, which gives a cross section of 0.4 pb; our signal (5) has a
combined cross section $\times$ branching ratio of
$$
  \sigma(pp\to t\bar{t}H) B^2(t\to bl^+\nu) B(H\to q\,\bar{q}\,or\,gg)
  = 0.4~\rm{pb} \times (2/9)^2 \times 0.95 = 18.8~\rm{fb}
\eqno(6)
$$
at the LHC. This is still significantly larger than the $H \to \gamma\gamma$
signal with
a single lepton tag.
Although a b-tagging efficiency of $50 \%$ is suggested to likely at the
TEVATRON  and at LHC, I assume only a $40 \%$ efficiency, and I assume a
lepton recognition efficiency of $90\%$, which has been achieved at LEP. 
To reduce most of the QCD background, I put a 15 GeV cut on the minimum
$p_T$ of each of the leptons and jets and place a 30 GeV minimum on the
total missing $E_T$. An isolation cut is then applied to each lepton,
requiring it to be at least $\sqrt{\Delta\eta^2 + \Delta\phi^2} = 0.4$
units away from any of the four jets. This has the effect of guaranteeing
that almost all remaining events are of the type $t\bar{t} + 2$ hard jets.
The same amount of separation is also applied to $e^+e^-$ and $\mu^+\mu^-$
pairs to avoid QED background.  

I now consider possible sources of background to the signal. There are
three major types of background events:

(1) Sources of background other than $t\bar t$ that produce $l^+ l^-$ pairs
are suppressed by at least two factors of either $\alpha_{weak}$ or
$\alpha_{em}$. 
These events do not have intrinsic missing
$E_T$ and will probably not pass the missing $E_T$ cut. Thus, non
$t\bar{t}$ contributions to the background are likely to be small.

(2) By far, the most important background is the QCD process
$$
  pp \to t{\bar t} + 2~\rm jets,
\eqno(7)
$$
where the jets come mostly from initial state radiation. Both the $t$ and
$\bar{t}$ decay, as before, semileptonically to give the required
signature.

Experience shows that $t\bar{t}$ pairs produced at the LHC are almost always
accompanied by extra high $p_T$ jets, and I make the ansatz that
there will be an average of 2.0 jets with $p_T > 15$ GeV in every $t\bar{t}$
event. By assuming a Poisson distribution for the number of extra jets, the
$t\bar{t}$ cross section with two and only two accompanying jets with $p_T >
15$ GeV is found to be 0.41 nb.  Folding in the leptonic branching ratio for
top quarks reduces this number to 0.02 nb.  The Poisson
distribution peaks at the average value, and I take this to be a
conservative estimate.

There is also a combinatorial factor for forming two-jet pairs from the four
jets in each event, and  this enhances the background by a factor of 6. Compared
to 18.8 fb for the signal, the background is larger by a factor of $2 \times
10^3$.  As seen later this will be suppressed severely by
additional cuts (see Table I).  In addition, the background events will not
have any special feature in the two-jet invariant mass spectrum while the
signal will show a prominent peak around the mass of the Higgs boson.

(3) In addition to $t\bar{t}H$, there are also $t\bar{t}W^{\pm}$
and $t\bar{t}Z$  events in which the $W^{\pm}$ and $Z$ decay to two jets,
thus producing the same signature. The production cross section for
$W^{\pm}$ is about one-tenth of that for the $Z$.  This together
with a smaller mass makes the $W^{\pm}$ events less important.
The $t\bar{t}Z$ signal is an exact analogue of $t\bar{t}H$ for $m_H$
close to the $Z$ mass. With $m_H$ near 114~ \GeVc2, I expect the two to have 
very similar cross
sections; however, they should have no significant overlap.  The estimated 
cross section for $t\bar{t}Z$ is roughly 0.3--0.4 pb for $m_t=175$. The
mass resolution for $Z\to 2$ jets at the LHC was estimated 
to be roughly ${}\pm 5$ GeV in a similar energy dependent
situation.

This signal not only turns out to be benign, it actually works to our
advantage. Unlike the Higgs, the $Z$ also decays to charged lepton pairs
with a large branching ratio (6\%). In my case, this gives a signature of
4 leptons and 2 jets, which has very little background if I further
require two of the leptons to reconstruct to a $Z$. Therefore, the
$t\bar{t}Z$ cross section can be measured by reconstructing the two-lepton
invariant mass in  the 4 leptons + 2 jets events.  I expect that most of
the theoretical uncertainty in the ratio of the two cross sections, $pp
\rightarrow t\bar{t} H$ and $pp \rightarrow t \bar t Z$,  cancels, and the
cross section for $t\bar{t}H$ can then be reliably inferred from that of
the $t\bar{t}Z$. 
Finally, this calibration from the $Z$ is absent in the $\gamma\gamma$
channel because $Z \to \gamma\gamma$ is forbidden on account of anomaly
cancellation.

\n\n
\noindent{\bf 3. SIMULATION}
\n
\par
Both the signal and background are simulated using Monte Carlo methods at
the parton level. As a preliminary study, simplified distributions in phase
space are used for each scattering and decay process. The total cross
sections are then normalized to published values.

For my signal, the parton cross section is given on purely dimensional
ground by 
$$
  d\sigma(gg \to t\bar{t}H) \propto \frac{1}{~\hat{s}^2}\,d\Phi_3~,
\eqno(8)
$$
where $d\Phi_n$ is the $n$-body Lorentz invariant phase space:
$$
  d\Phi_n = \delta^{(4)}(P - {\textstyle\sum} p_i)\,
	    \prod_{i=1}^n \frac{d^3p_i}{2E_i}~,
\eqno(9)
$$
where $\hat{s}$ is the center of mass energy of the partons and $P$ is
their total momentum.  Each of $t,\ \bar{t}$ and $H$ is then decayed
independently to $b\,l^+\nu$, $\bar{b}\,l^-\bar{\nu}$ and $b \bar{b}$
respectively to give us the signature of 4 jets + 2 leptons + missing
$E_T$. I take $m_t = 175,\ m_H = 114~\GeVc2$ and 0.4 pb for the total cross
section.

For the background, I first take
$$
  d\sigma(gg \to t\bar{t}gg) \propto \frac{1}{~\hat{s}^3}\,d\Phi_4~,
\eqno(10)
$$
and normalize it to the $t\bar{t}gg$ cross section. Just as
in the case of $t\bar{t}H$, I assume that the dominant contribution to the
$t\bar{t}$ cross section comes from gluon fusion.  Since the top quarks
are heavy and hardly radiate, I next assume that the two extra jets are
gluons coming from initial state radiation only. The angular and energy
distributions of the gluons, in addition to Eq.~(10), are assumed to follow
the Altarelli-Parisi function
$$
  P_{gg}(z) = c_{gg} \left( \frac{z}{1-z} + \frac{1-z}{z} + z(1-z) \right),
\eqno(11)
$$
where $z$ is the fractional momentum of the initial gluon after radiation:
$$
  z = \frac{E_i - {p_L}_i}{E_{i-1} - {p_L}_{i-1}}~,
\eqno(12)
$$
where the subscripts $i-1$ and $i$ refer to the initial gluon before and
after radiation. I also use the approximation where the further splitting of a radiated 
virtual gluon is replaced by two sequential radiations from the same initial gluon. 
The $t$ and $\bar{t}$ are decayed as before to give us the required signature.

To mimic a real experimental situation, the energy of each parton is smeared
to reproduce the proposed detector resolution for jets and leptons,
and I impose an isolation cut on each of the hadronic partons (jets) in
$\eta$-$\phi$ space so that no two jets are within a distance of 0.5 units
from each other:  
$$
  \Delta R = \sqrt{\Delta\eta^2 + \Delta\phi^2} > 0.5~.
$$
A rapidity cut of $\eta<2.5$ is also applied to each of the jets and leptons.
\n\n
\noindent{\bf 4. RESULTS}
\n
\par
I generate the number of events for the signal and background that
corresponds to an integrated luminosity of 100 fb$^{-1}$ with an 
LHC luminosity of
$10^{33}~\rm cm^{-2}s^{-1}$. The following cuts are
then performed on both types of events: 
(1) A 15 GeV minimum $p_T$ cut is first applied to all jets and leptons. 
(2) A second $p_T$ cut is then applied to the four hadronic jets:
For each of the six pairs of jets, I form the scalar sum of the two
individual $p_T$ and require it to be larger than 60 GeV\null. This will
severely suppresses the background while keeping the signal almost
unchanged.  (3) To try to guarantee that the two leptons originate from top
quark decays, I require each lepton to be at least a distance  $\Delta R =
0.4$ away from any jet in $\eta$-$\phi$ space. The two leptons are required
to be  separated by the same amount. (4) The missing transverse energy in
each event must be larger than 30 GeV. 

There are many less important sources of background such as $Z+{\rm 4~jets}
+{\rm missing}~E_T$ and $b\bar{b} + 2g$, etc.  To ensure that they do not
significantly affect our result, I also apply the following cut:
The invariant mass of all the observed
particles should be larger than $2m_t + m_H$.
This greatly enhancing the probability of having
heavy particles in the final state.
The results of all the cuts are shown in Table~I in the order of their
applications. 

$$\vbox{
  \halign{\quad#\hfil & \quad\hfil#\hfil & \quad\hfil#\hfil \cr
  {\bf \hfill Table I.} &&\cr
  \noalign{\vskip1ex\hrule \vskip1pt \hrule \vskip1ex}
	      & Signal ($t\bar{t}H$)      & Background ($t\bar{t}gg$) \cr
  \qquad Cuts & $100\% \approx 18.8$ fb & $100\% \approx 2.0$ pb \cr
  \noalign{\vskip1ex \hrule \vskip1ex}
  All $p_T > 15$ GeV                      &  51\%  &  64\% \cr
  All  $p_{T1}+p_{T2} > 60$ GeV           &  49\%  &  11\% \cr
  Missing $E_T > 40$ GeV                  &  48\%  &  11\% \cr
  Observed invariant mass $>$ 464 GeV     &  33\%  &  3.2\% \cr
  Isolation cuts ($j$-$j$, $l$-$j$ and $l$-$l$) &  17\%  & ~1.3\% \cr 
  \noalign{\vskip1ex \hrule}}
}$$

To extract the signal from this background, I need to reconstruct the
Higgs mass from two-jet pairs. Figure~1a shows the distribution of the
two-jet invariant mass of the signal and the $Z$ peak. It has a prominent peak about
$m_H$ = 114~\GeVc2\ containing 208 counts  in the 8 \GeVc2\ or so region under the peak.  The
width of the resonance matches roughly the two-jet resolution of 8 GeV
achievable at the LHC.  The $t\bar{t}gg$ background is shown in Fig.~1b.
Over the same 8 GeV range under the Higgs peak, the background contains
roughly 1680 counts.  Thus, I obtain a significance, ${\rm
signal}/\sqrt{{\rm background}}=5.1$.
The combined result is shown in Fig.~1c.  This figure shows with
 our approximations a recognizable  signal
over the QCD background despite the fact that the cuts applied have not been
fully optimized. In view of all the approximations I have made, the
background in the region of interest can easily be off by a factor of two or
three.
For example, the Altarelli-Parisi splitting function I used in Eq.~(11)
tends to generate gluons softer and closer to the beam line than would occur
in real background events.
However, I have been  generous in normalizing the overall
cross section for the background.  If more optimal cuts are also used, I
believe the signal could  still be measured over a slightly higher background.
I believe
that reconstruction from jets is a promising technique in the detection of the
Higgs boson, and, used in conjunction with related methods \Ref{8--9}, it should give
definitive evidence for Higgs detection.
QCD NLO corrections \Ref{6} to the process $g\,g \rightarrow H$ indicate
a $K$ factor of $\approx 2.2$, and this could improve the signal
recognition. Although jet smearing has been used to give a more realistic
representation of the jet and lepton signals, the actual fragmentation of
jets is hard to simulate \Ref{9}, and this could cancel the NLO effect.
Finally, these results can be used in conjunction with other signals, and
this should lead to Higgs detection within the first two years of LHC
operation.

The author has benefited from the use of the Queen's University Belfast and
Dublin University 
parallel processor.
\n\n
\noindent{\bf REFERENCES}
\n
\item{[1]} Particle data group, Euro. Phys. Jr. \ B {\bf 15}, 103, 274 (2000). 

\item{[2]} L3 Collaboration, /hep-ex/0011043, Phys. \ Lett.\ B {\bf 495}, 18-25(2000). 

\item{[3]} ALEPH Collaboration, /hep-ex/00111045, Phys. \ Lett.\ B {\bf 495}, 1-17(2000). 

\item{[4]} Z.~Kunszt, Nucl.\ Phys.\ {\bf B247}, 339 (1984);  D.~Dicus and
S.~Willenbrock, Phys.\ Rev.\ D {\bf 39}, 751 (1989).

\item{[5]} T.~Garavaglia, Waikwok Kwong, Dan-Di Wu, Phys.\ Rev.\ D 
{\bf 48}, R1899-R1903 (1993). 

\item{[6]} M.~Spira, A.~Djouadi, D.~Graudenz, and P.~M.~Zerwas, Nucl. \
Phys. {\bf B 453}, 17 (1995).

\item{[7]} M.~Spira, Fortsch. \ Phys. {\bf 46}, 203 (1998).

\item{[8]} D. Denegri, {\it The CMS Detector and Physics at LHC}, CERN-PPE/95-183.

\item{[9]} D.~Froidevaux, E.~ Richter-Was, {\it Is the $H\rightarrow b \bar
b$ observable at LHC}, Atlas Internal Note, Phys-NO-043 (1994).
\n\n

\noindent{\bf FIGURE CAPTION}
\n\n
FIG.~1. Two-jet invariant mass distributions. The vertical axes shows
counts/2 \GeVc2 bin. The horizontal axes shows two-jet invariant mass 
values in \GeVc2.

(a) $t\bar{t}H$ signal and $t\bar{t}Z$ events. 

(b) QCD $t\bar t gg$ background. 

(c) Signal plus background.
\vfil\eject  
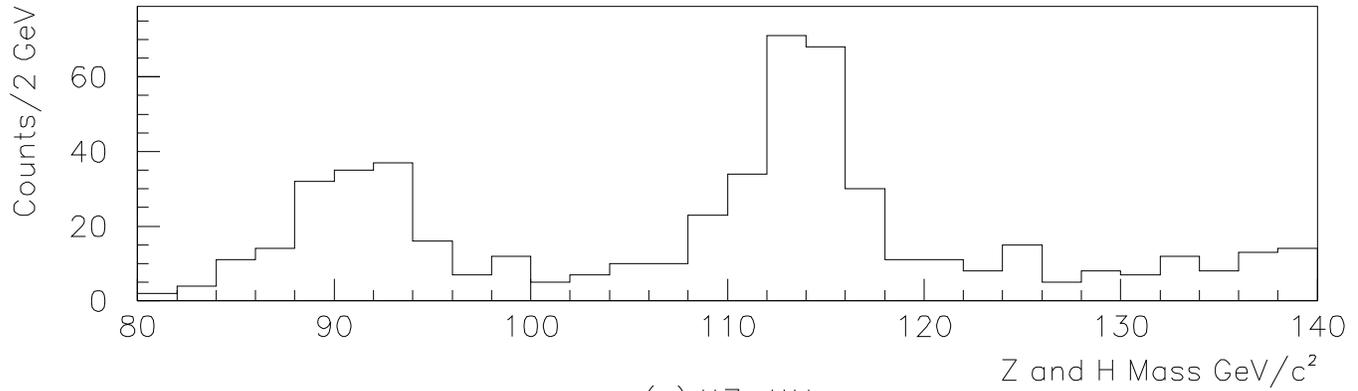
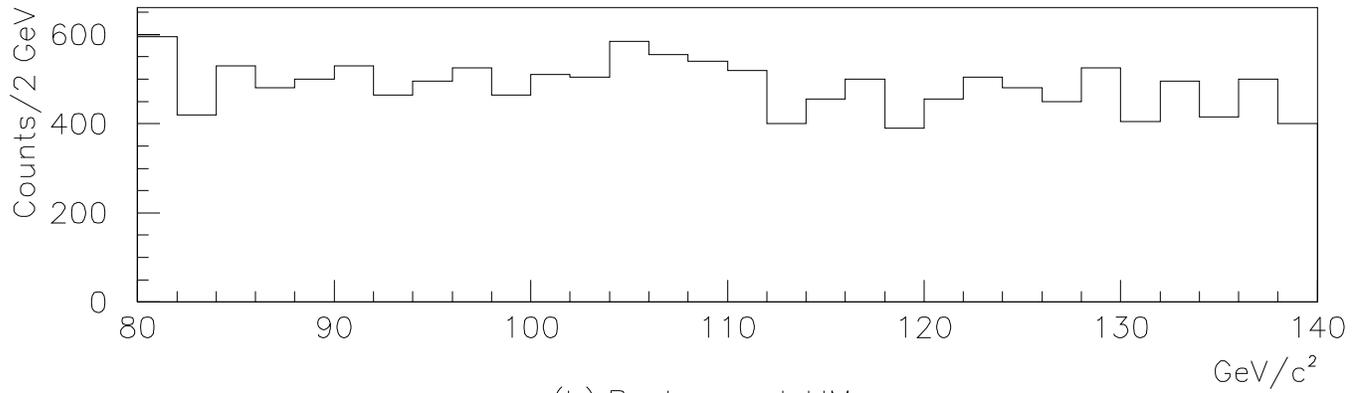
\bye